\documentstyle[12pt]{article}

\input epsf.sty
\catcode`@=11

\def\beqra{\begin{eqnarray}} \def\eeqra{\end{eqnarray}}
\def\beqast{\begin{eqnarray*}} \def\eeqast{\end{eqnarray*}}
\def\beq{\begin{equation}}	\def\eeq{\end{equation}}
\def\be{\begin{enumerate}}   \def\ee{\end{enumerate}}

\def\fnote#1#2{\begingroup\def\thefootnote{#1}\footnote{#2}\addtocounter
{footnote}{-1}\endgroup}

\def\sppt{Research supported in part by the
Robert A. Welch Foundation and NSF Grant PHY 9009850}

\def\utgp{\it Theory Group,  Department of Physics \\ University of Texas,
 Austin, Texas 78712}

\def\la{\lambda}

\def\si{\sigma}

\def\del{\delta}
\def\Del{\Delta}

\def\om{\omega}

\def\mn{\mu\nu}

\def\til{\tilde}
\def\rta{\rightarrow}
\def\eqv{\equiv}

\def\pa{\partial}

\def\lag{\langle}
\def\rag{\rangle}

\def\ch{\@startsection{section}{1}{\z@}{-3ex plus-1ex minus-.2ex}%
	{2ex plus.2ex}{\large\sc{\normalsize}}}

\def\cl{{\cal L}}
\def\cm{{\cal M}}

\def\cO{{\cal O}}

\def\raisenot{\raise .5mm\hbox{/}}
\def\nota{\ \hbox{{$a$}\kern-.49em\hbox{/}}}
\def\notA{\hbox{{$A$}\kern-.54em\hbox{\raisenot}}}
\def\notb{\ \hbox{{$b$}\kern-.47em\hbox{/}}}
\def\notB{\ \hbox{{$B$}\kern-.60em\hbox{\raisenot}}}
\def\notc{\ \hbox{{$c$}\kern-.45em\hbox{/}}}
\def\notd{\ \hbox{{$d$}\kern-.53em\hbox{/}}}
\def\notbd{\ \hbox{{$D$}\kern-.61em\hbox{\raisenot}}} 
\def\note{\ \hbox{{$e$}\kern-.47em\hbox{/}}}
\def\notk{\ \hbox{{$k$}\kern-.51em\hbox{/}}}
\def\notp{\ \hbox{{$p$}\kern-.43em\hbox{/}}}
\def\notq{\ \hbox{{$q$}\kern-.47em\hbox{/}}}
\def\notW{\ \hbox{{$W$}\kern-.75em\hbox{\raisenot}}}
\def\notz{\ \hbox{{$Z$}\kern-.61em\hbox{\raisenot}}}

\def\notpa{\hbox{{$\partial$}\kern-.54em\hbox{\raisenot}}}

\def\fo{\hbox{{1}\kern-.25em\hbox{l}}}  

\def\tr{{\rm Tr}}

\def\and{\,{\rm and}\,}

\def\7#1#2{\mathop{\null#2}\limits^{#1}}	
\def\5#1#2{\mathop{\null#2}\limits_{#1}}	

\def\inbar{\vrule height1.5ex width.4pt depth0pt}
\def\IB{\relax{\rm I\kern-.18em B}}
\def\IC{\relax\leavevmode\hbox{\,$\inbar\kern-.3em{\rm C}$}}
\def\ID{\relax{\rm I\kern-.18em D}}
\def\IE{\relax{\rm I\kern-.18em E}}
\def\IF{\relax{\rm I\kern-.18em F}}
\def\IG{\relax\leavevmode\hbox{\,$\inbar\kern-.3em{\rm G}$}}
\def\IH{\relax{\rm I\kern-.18em H}}
\def\II{\relax{\rm I\kern-.18em I}}
\def\IK{\relax{\rm I\kern-.18em K}}
\def\IL{\relax{\rm I\kern-.18em L}}
\def\IM{\relax{\rm I\kern-.18em M}}
\def\IN{\relax{\rm I\kern-.18em N}}
\def\IO{\relax\leavevmode\hbox{\,$\inbar\kern-.3em{\rm O}$}}
\def\IP{\relax{\rm I\kern-.18em P}}
\def\IQ{\relax\leavevmode\hbox{\,$\inbar\kern-.3em{\rm Q}$}}
\def\IR{\relax{\rm I\kern-.18em R}}
\def\sed{\hbox{{\sf S}\kern-.4em\hbox{\sf S}}}
\def\ZZ{\relax{\sf Z\kern-.4em Z}}
\def\zed{\hbox{{\sf Z}\kern-.4em\hbox{\sf Z}}}
\def\smIR{\hbox{{\footnotesize\rm I}\kern-.2em\hbox{\footnotesize\rm R}}}
\def\smIO{\ \hbox{{\footnotesize\rm I}\kern-.4em\hbox{\footnotesize\bf O}}}
\def\smIQ{\ \hbox{{\footnotesize\rm I}\kern-.5em\hbox{\footnotesize\bf Q}}}
\def\IGa{\relax{\rm I}\kern-.18em\Gamma}
\def\IPi{\relax{\rm I}\kern-.18em\Pi}
\def\IQt{\relax\leavevmode\hbox{$\kern.3em\inbar\kern-.3em\Theta$}}
\def\IOm{\relax\hbox{$\kern3.48pt\inbar\kern1.8pt\inbar\kern-5.28pt\Omega$}}

\def\ca#1{\relax\ifmmode {\cal#1} \else$\cal#1$\fi}	
\def\Sf#1{\relax\ifmmode\hbox{\sf#1}\else{\sf#1}\fi}	
\def\fibby{\ifcase\@ptsize 			
		\font\tenrm=cmfib8\or		
		\font\elvrm=cmfib8 scaled\magstephalf\or	
		\font\twlrm=cmfib8 scaled\magstep1 \fi}		
\def\TeXey{\ifcase\@ptsize\or\or 		
		\font\twlrm=cmr10 scaled\magstep1	
		\font\twlmi=cmmi10 scaled\magstep1	
		\font\twlit=cmti10 scaled\magstep1	
		\font\twlbf=cmbx10 scaled\magstep1\fi}	

\def\ch{\@startsection{section}{1}{\z@}{-3ex plus-1ex minus-.2ex}%
	{2ex plus.2ex}{\large\sc}}
\def\sch{\@startsection{subsection}{2}{\z@}{-1.5ex plus-1ex minus-.2ex}%
	{1pt plus.2ex}{\sc}}
\def\ssch{\@startsection{subsubsection}{3}{\z@}{-1ex plus-1ex minus-.2ex}%
     	{1pt plus.2ex}{\small\sc}}
\def\seceq{\@addtoreset{equation}{section}
	\def\theequation{\thesection.\arabic{equation}}}	

\def\con{\ifmmode \hbox{\bf*} \else{\bf*}\fi}	
\def\scon{\ifmmode \hbox{\footnotesize\rm\bf*} \else{\footnotesize\rm\bf*}\fi}

\def\0#1{\relax\ifmmode\mathaccent"7017{#1}
	\else\accent23#1\relax\fi}		

\def\haf{\frac{1}{2}}

\def\place#1#2#3{\vbox to0pt{\kern-\parskip\kern-7pt
                             \kern-#2truein\hbox{\kern#1truein #3}
                             \vss}\nointerlineskip}

\def\illustration #1 by #2 (#3){\vbox to #2{\hrule width #1 height 0pt depth
0pt
                                       \vfill\special{illustration #3}}}

\def\scaledillustration #1 by #2 (#3 scaled #4){{\dimen0=#1 \dimen1=#2
           \divide\dimen0 by 1000 \multiply\dimen0 by #4
            \divide\dimen1 by 1000 \multiply\dimen1 by #4
            \illustration \dimen0 by \dimen1 (#3 scaled #4)}}

\def\ijmp#1#2#3{{\it Int. J. Mod. Phys.} {\bf A#1} (19#2) #3 }

\def\npb#1#2#3{{\it Nucl. Phys.} {\bf B#1} (19#2) #3 }

\def\prd#1#2#3{{\it Phys. Rev.} {\bf D#1} (19#2) #3 }
\def\pr#1#2#3{{\it Phys. Rev.} {\bf #1} (19#2) #3 }

\def\prl#1#2#3{{\it Phys. Rev. Lett.} {\bf #1}(19#2) #3 }

\mark{{}{}}   

\def\ps@headings{\let\@mkboth\markboth
\def\@oddfoot{}\def\@evenfoot{}
\def\@evenhead{  \sl \leftmark \hfil}
\def\@oddhead{\hbox{}\hfil \sl \rightmark   }
}

\def\ps@myheadings{\let\@mkboth\@gobbletwo
\def\@oddhead{\hbox{}\sl\rightmark \hfil \rm\thepage}%
\def\@oddfoot{}\def\@evenhead{\rm \thepage\hfil\sl\leftmark\hbox {}}%
\def\@evenfoot{}\def\chaptermark##1{}\def\sectionmark##1{}%
\def\subsectionmark##1{}}

\catcode`@=12
\seceq
\begin{document}

\rightline{UTTG-13-94}
\rightline{\today}

\vspace{24pt}
\begin{center}
\large{\bf Non-Perturbative Decoupling of Heavy Fermions}

\normalsize

\vspace{36pt}
Moshe Rozali\fnote{\dag}{\sppt}

\vspace{8pt}
\utgp
\end{center}

\abstract
{We show that heavy fermions decouple from the low energy physics also in
non-perturbative instanton effects. Provided the heavy fermions are lighter
than the symmetry breaking scale, all the instanton effects should be expressed
as local operators in the effective Lagrangian.  The effective theory itself
doesn't admit instantons.  We present the mechanism which suppresses instantons
in the effective theory.}

\vfill
\pagebreak
\baselineskip=21pt
\section{Introduction}

\indent\indent
In a recent paper \cite{BD} non-perturbative instanton effects were claimed to
violate decoupling.  Several solutions to the problem were suggested
\cite{{G},{Hsu}}.  In this paper we reformulate the problem in the context of
the standard model and clarify the the process of integrating out heavy
fermions.

In what follows we consider the standard model with three families and
inter-family mixing.  The color and hypercharge indices are suppressed, and the
lepton doublets are left undisplayed, since they are irrelevant for the
discussion.

The Lagrangian of the model is (see the appendix for notations):

\beq
\cl=-\haf\tr(W_{\mn}W^{\mn})+\haf(\pa_\mu\til\si)^2-\frac{\la}{4}
\,(\til\si^2-v^2)^2+\frac{\til\si}{4}\,\tr(D_\mu U^{\dagger} D^\mu U) +L_{\rm
fermionic}\,.
\eeq

For one quark doublet the fermion action is:
\beq
\cl_\psi=i\psi^{\dagger}
\si^\mu D_\mu\psi+i\bar\psi^{\dagger}\si^\mu\pa_\mu\bar\psi-\left(\bar\psi^T
\frac{\cm_{q}}{v}\,\til\si U\psi+{\rm c.c.}\right)
\eeq
We assume equality of the top and bottom mass, $\cm_q=m_t\fo$, unless otherwise
indicated. $\lag\til\si\rag=\si$ breaks the symmetry,  $\si\eqv\til\si-v$
is the physical Higgs. $U$ is the angular Higgs (the would-be Goldstone
bosons).

In theories with Higgs fields no exact instanton solutions exist.  However,
there exist approximate solutions, known as ``constrained instantons''
\cite{AF}.  In the next section we review the constrained instanton
construction and the resulting 'tHooft operator \cite{'tH}.  This operator
violates baryon number by 3 units.

We integrate out one heavy quark doublet, assuming $m_\si$ and $m_W$ are
lighter than $m_t$, so the effective theory contains the same bosonic fields.
After integrating out the fermions the theory's predictions for low energy
processes should be obtainable from a local effective Lagrangian --- this is
what is meant by ``decoupling'' in this context.

First, the effective Lagrangian should reproduce the original 't Hooft
operator.  This matching is shown in section 3.  Secondly, it must predict
vanishing amplitudes for $\Del B=2$ processes, which would naively appear as
't Hooft operators in a two family model.  This is shown in detail in Section
4.  We present our conclusions in Section 5.

\section{The Constrained Instanton:}

\indent\indent In this section we review instanton effects in the standard
model.  We follow some of the notation of \cite{ES}, and the reader is referred
to the latter for more details.

Fermion number violating amplitudes arise in a semi-classical expansion around
an Euclidean configuration with a non-vanishing winding number.  Schematically
one has the following Euclidean Lagrangian:
\beq
\cl_E=\psi_A\,M(A)\psi_B + B(A)
\eeq
$\psi_A,\psi_B$ are Euclidean fermions \cite{ES}, and $A$ some bosonic fields.
Expanding to leading $\hbar$ order around some configuration $A_0$:
\beq
\cl_E=\psi_A\,M(A_0)\psi_B+ B(A_0) + {\rm corrections}\;.
\eeq
The minimal non-vanishing amplitude is then:
\beq
\lag\psi_1(x_1)\ldots\psi_r(x_r)\rag =C
e^{-S_0}\psi_0^1(x_1)\ldots\psi_0^r(x_r)\,.
\eeq

$C$ is a constant calculated from the bosonic fluctuations around
$A_0~,~S_0=\int d^4 x B(A_0)$ has to be finite to contribute to the path
integral.  $r$ is the number of zero modes of the operator $M(A_0)$, related to
chiral anomalies by index theorems.

In the case considered here, the relevant configuration is the constrained
instanton \cite{AF}, which is an approximate saddle point with a finite
action.  After an Euclidean rotation and rescaling, the bosonic action is:
\beq
\cl_B=\frac{1}{g^2}\,\left[\haf\,\tr(W_{\mn}W_{\mn})+\kappa^2\tr(D_\nu
M^{\dagger} D_\nu M)+ \kappa^2 (\til\si^2-m_\si^2)^2\right]
\eeq

$M=\til\si\,U$ is the Higgs Field.  $\si=\til\si-m_\si$ is the physical Higgs.
The semi-classical limit is $g^2$ small, $m_\si^2$ and
$\kappa=\frac{m\om}{m\si}$ fixed.

The Euclidean equations of motion are:
\beqra
&& D_\mu W_{\mn}+ \frac{i\kappa^2}{2}\,(D_\nu M)^{\dagger}M-
\frac{i\kappa^2}{2}\,M^{\dagger} D_\nu M=0
\nonumber \\
&& D_\mu^{\dagger}D_\mu M+\left(\haf\,\tr(M^{\dagger}M)-m_\si^2\right)M=0
\eeqra
with no source terms  $(\la=0$ in (1.1)) there's a finite action
solution:
\beqra
W_\mu^0&=& \frac{2\rho^2}{x^2(x^2+\rho^2)}\,x_\nu\bar\tau_{\mn} \nonumber \\
M^0 &=& \left(\frac{x^2}{x^2+\rho^2}\right)^{1/2}\;m_\si\fo
\eeqra

However, the full action (2.4) is infinite for this configuration.  Therefore
we
try to deform (2.6) to get another finite action solution.  Try:
\beqra
M &=& M^0+\del M \nonumber \\
W_\mu &=& W_\mu^0+\del\,W_\mu
\eeqra

We get equations of the form
\beq
\cO\del\phi=J\,.
\eeq
  Where $\phi$ is either $W_\mu$ or $M,~\cO$ is some differential operator.
$\del\phi$ is also subjected to some boundary conditions to ensure a finite
action (namely $\del\phi\rta 0$ fast enough at infinity).

The two above conditions on $\del\phi$ are inconsistent.  The operator $\cO$
has
zero modes, so the source $J$ propagates to infinity and determine the boundary
values of $\del\phi$.  These boundary values correspond to the required ones
only if $J$ is perpendicular to the zero modes of $\cO$.

Resolution of this problem was given in \cite{AF}.  Constraining the path
integral in a particular way has the effect of adding operators to the
Lagrangian.  The modified Lagrangian is called the constrained Lagrangian.  The
new operators are required to vanish fast enough at infinity (faster than
$\frac{1}{|x|^4}$ for the configuration (2.6)) so they don't change the
finite action boundary conditions.  However, they do change the source $J$ in
(2.8), and can be tuned to make it perpendicular to the zero modes.  Therefore,
one can get instanton solutions to the constrained Lagrangian, which are
approximate solutions to the original Lagrangian.

Asymptotic expressions for the resulting configuration can be obtained in two
different regions \cite{{AF},{ES}}.  First, in the unbroken phase
($|x|< v^{-1}$) we expect to get approximately an 'tHooft instanton of size
$\rho$, which solves the equations:
\beqra
D_\mu W_{\mn} = 0 \nonumber \\
D^2 M=0
\eeqra
These equations cease to be a good approximation to the full equations of
motion where the neglected terms start to dominate. This determines the range
of
validity of (2.9).  One gets:
\beqra
W_\mu &=& \frac{2\rho^2}{x^2(x^2+\rho^2)}\;x_\nu\bar\tau_{\mn}+\cdots
\qquad\qquad ~~|x|\ll m_\om^{-1} \nonumber \\
M &=& \left(\frac{x^2}{x^2+\rho^2}\right)^{1/2}\, m_\si\,\fo+\cdots
\qquad\qquad ~~|x|\ll m_\si^{-1}
\eeqra
To get an asymptotic expression for $|x|\gg\rho$ one leaves in the equations of
motion just the leading terms at infinity (For the configurations (2.10))
\cite{AF}.  All other terms in (2.5) are replaced by a delta function source.
The resulting equations are:
\beqra
&& \pa^2M - m_\si^2 \,M\,\propto\,\del^{(4)}(x) \nonumber \\
&& \pa_\mu W_{\mu\nu} + i\kappa^2 m_\si^2\,W_\nu \,\propto\,\del^{(4)} (x)\;.
\eeqra

The terms in (2.11) vanish as $\frac{1}{|x|^4}$ as $|x|\rightarrow\infty$,
whereas all the other operators are suppressed by powers of
$\frac{\rho^2}{|x|^2}$, including the terms in the constrained Lagrangian added
to impose the constraints.

This gives the asymptotic expansion:
\beqra
W_\mu &=& -\rho^2\bar\tau_{\mn}\pa_\nu\, G(m_\om,|x|)+\cdots  \nonumber \\
M &=& m_\si\,\fo\,\left(1-\haf\,\rho^2G(m_\si,|x|)+\cdots\right)\qquad
|x|\gg\rho
\eeqra
where
\beqra
&& G(m,|x|) = m^2\;\frac{K_1(m|x|)}{m|x|} \nonumber \\
&& (\Box-m^2)\,G\,(m,|x|)=-4\pi^2\,\del(x)\,.
\eeqra
One assumes $\rho v\ll1$, large instantons cannot be treated that way to yield
an approximate saddle point.  The above expressions are first order in the
small parameter $\rho v$.  Since $\rho\ll v^{-1}$ the regions of the above
expressions are overlapping, so the configuration is well defined everywhere.
The two expressions have to be patched in the intermediate region
$(\rho<|x|<v^{-1}$).

The fermionic zero modes of this configuration have similar expansions
\cite{ES}.  It is also assumed that $\cm_q<v$, which is the case considered in
this paper.

\section{The low energy effective Lagrangian:}

\indent\indent Consider the energy range $m_\om<E<\,m_t<v$.  In this range
all virtual configurations smaller than $m_t^{-1}$ should be expressed as a
series of local operators.  This includes perturbative configurations (like
heavy quark loops) as well as small instantons.  Therefore the 't Hooft
operator should be included in the effective Lagrangian, as a part of the
matching process.

The 't Hooft operator involves necessarily a heavy quark.  Therefore its not
part of the physics to be described by the low energy effective theory.
However, with family mixing there are low energy baryon number violating
operators, resulting from virtual heavy quarks.  An example of a dimension 21
operator is given in Figure 1.  This operator and similar ones should be a
part of the effective Lagrangian.

\vskip 15pt
\centerline{
\epsfxsize=5.2in
\epsfbox{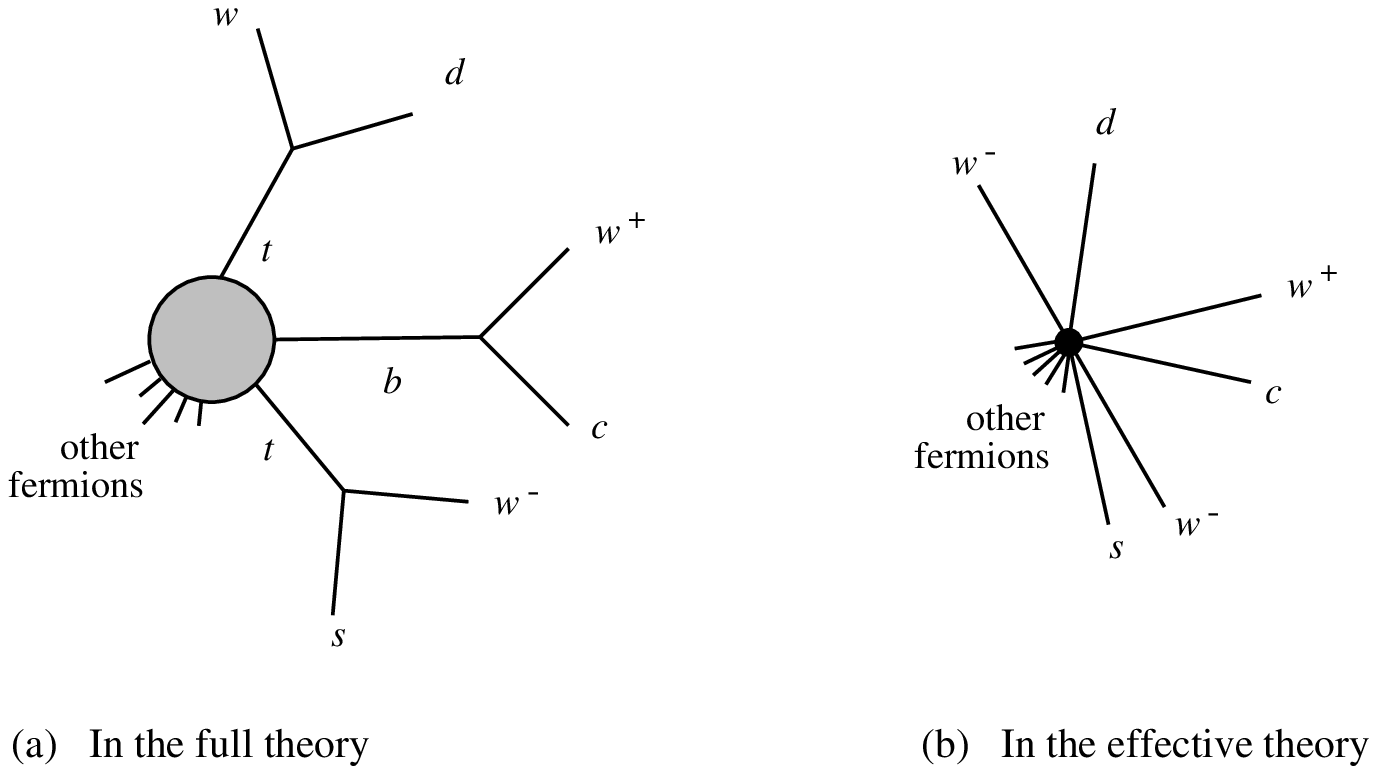}}

\vspace{12pt}

\centerline{Figure 1}

\vspace{12pt}
The situation is similar in the case of integrating out just the top (assuming
$m_t\gg m_b)$.  The resulting theory is non-linearly realized \cite{{C},{FMM}}.
Some of the fermions transform linearly under SU(2), and the bottom transforms
non-linearly, as a part of an incomplete multiplet.  To build SU(2) invariants
it's convenient to use the composite field $U\left(0\atop b\right)$ which
transform linearly as an SU(2) doublet \cite{F}.

Expanding $U=\exp\left(i\frac{\zeta^a\tau^a}{v}\right)$ shows that replacing
the complete doublet by $U\left(0\atop b\right)$ in the 't Hooft operator
recovers the original operator with $b$ emitted, but also a series of operators
involving $\zeta^a$ (the longtitudal gauge bosons).  These additional
operators are suppressed by powers of $v^{-1}$, and represent effects of
integrating out the top.

\section{Instanton Suppression in the effective \newline theory:}

\indent\indent
An instanton solution to the low energy theory, and corresponding fermionic
zero modes for the remaining two quark doublets would contradict the full
theory's predictions and therefore violate decoupling.  We show now that the
low energy theory doesn't admit instantons.

Integrating out heavy fermions induces a series of higher dimensional terms in
the effective Lagrangian, see for example \cite{{FD},{F}}.  It was suggested
\cite{G} that some of these operators give rise to infinite action (zero
measure) for the original constrained instanton.  However, the divergences come
from the short distance behavior of these operators, where the effective
action cannot be used.It is also conceivable that one can modify the
original solution (by changing the constraints, for instance) to another
approximate solution which accommodates the higher dimensional terms and has a
finite action.  This would represent threshold corrections to the instanton
coming from integrating out the fermions.

In the following we demonstrate that no such solution is possible.  We assume,
ex absurdo, that the constrained Lagrangian can be adjusted to enable finite
(effective) action boundary conditions.  We study the resulting asymptotic
expression for $|x|\gg \rho$ and conclude that it cannot be patched to any
configuration with a non-zero winding number, thus contradicting the
assumption.

Assume then that we have an instanton configuration (2.10) at short distances.
We now try to get an asymptotic expression similar to (2.12).  Most new
operators in the effective theory will be replaced by a delta function source
in
(2.11), but some will also contribute to the left hand side of equation (2.11).
Consider for example the operator:
\beq
\Delta\cl=\frac{1}{4\pi^2m_\si}\,\si\tr\;(W_{\mn}\;W_{\mn})
\eeq
coming from a diagram shown in figure (2).

\centerline{
\epsfxsize=1.2in
\epsfbox{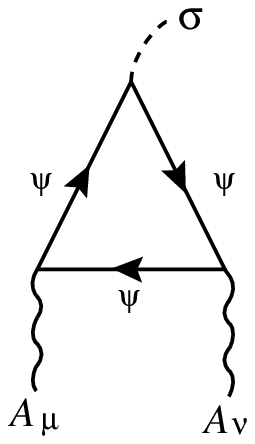}}

\centerline{Figure 2}

\vspace{12pt}
This operator cannot be ommited in the region described by equation (2.11),
since it dominates the
existing terms for $|x| < \frac{\rho}{\sqrt{\rho v}} $. In particular we get
the modified equation :
\beq
\kappa^2(\Box-m^2_\si)\si+\frac{1}{4\pi^2m_{\si}}\,\tr\,(W_{\mn}W_{\mn})
=2\pi^2\rho^2\,\del(x)\;.
\eeq

All other higher dimensional operators in the equations of motion are
suppressed by powers of $\frac{\rho^2}{|x|^2}$ or $(m_t|x|)^{-1}$.  Therefore
equation (4.2) holds for $|x|\gg m_t^{-1}$ (not for all $|x|\gg\rho$ as in
section 2).

Equation (4.2) is an equation of the form (2.8), which in general imposes
restrictions (orthogonality conditions) on the source.  To see these
restrictions multiply (4.2) by $G(m_\si,|x|$) and integrate from $r>m_t^{-1}$
to infinity.  The left hand side picks up just boundary terms at $r$.
Choosing $r<m_\si^{-1}$ and $r<m_\om^{-1}$, the unbroken solution (2.10) still
holds.  Therefore:
\beq
\frac{1}{v}\, \int_r^\infty \!\!d^4x\, \tr(W_{\mn}\,W_{\mn})\,
G(m_\si,|x|)\sim 0 \left(v\;\frac{\rho^2}{r^2}\right)\,.
\eeq
(Using:  $G(m_\si,|x|)\sim \frac{1}{|x|^2}+0(\rho v)$~) \\
or:
\beq
W_{\mu\nu}\,(|x|=r)\ll 0\left( \frac{\rho v}{r^2}\right)\,.
\eeq

This statement is modified by corrections to (2.10), (4.3) etc.  However, all
these corrections are of order $\rho v$.  To first order in $\rho v$, (4.4)
implies that $W_{\mu\nu}=0$ for $m_t^{-1}<r< m_\om^{-1}$.  This is inconsistent
with the assumption that we have an instanton in the unbroken phase.

The corrections mentioned above make $W_{\mn}\sim 0(\rho v)$.  This would give
winding number of order $\rho v$.  However, since $\rho v\ll 1$, this is still
inconsistent with having an instanton configuration, small corrections cannot
generate a winging number.  Therefore we conclude that there are no possible
approximate instanton solutions in the effective theory.

\section{Conclusions}

In this paper we have considered small instantons coupled to fermions lighter
than the symmetry breaking scale.  We have integrated out one heavy quark
doublet.  In this limit we have shown that there are no instantons in the low
energy theory.  The result came from considering the region where an instanton
is patched to an asymptotic expression.  Integrating out the fermions changes
the
asymptotic expression, making it impossible to patch it to an instanton.  Even
though the core of the instanton is not describable by the effective theory,
our considerations relied only on regions where the effective theory is valid.
This was possible due to the existence of an operator in the effective
theory that is not suppressed by powers of the heavy fermion mass, and
therefore can affect the region describable by the effective theory.

\section*{Acknowledgements}

The author wishes to thank W. Fischler for suggesting the problem, for his
invaluable advice and criticism and for his comments on the manuscript.  Useful
conversations with J. Distler, J. Feinberg, S. Paban and J.M. Rodriguez are
gratefully acknowledged. The author also thanks T. Gould for a useful
correspondence.

\section*{Appendix}
\def\theequation{A.\arabic{equation}}
\setcounter{equation}{0}

The standard model action can be obtained from the general theory of non-linear
realization \cite{C}, see for example \cite{F}.

The bosonic Lagrangian is
\beq
\cl_B=-\haf \,\tr(\hat W_{\mn}\,W^{\mn}+\hat B_{\mn}\hat
B^{\mn})+\haf\;(D_\mu M^{\dagger}\,D^\mu\, M)+ V\,(M^{\dagger}M)
\eeq
$W_\mu,B_\mu$ are SU(2) matrices
\beqra
\hat W_\mu &=& W_\mu^at^a \nonumber \\
\hat B_\mu &=& B_\mu t^3
\eeqra
$M=\til\si U$, the covariant derivatives are:
\beqra
D_\mu\til\si &=& \pa_\mu\til\si \nonumber \\
D_\mu\hat U& = & \pa_\mu\hat U+ig\hat W_\mu\hat U-ig'\hat U\hat B_\mu\;.
\eeqra
The top-bottom Lagrangian is:
\beq
\cl_\psi=i\psi^{\dagger}\si^\mu D_\mu\psi + i\bar\psi^{\dagger}\si^\mu
D_\mu\bar\psi-\left[\bar\psi^T\;\frac{M_q}{v}\,M\,\psi+\mbox{c.c.}\right]
\eeq

All the Fermions used are left handed Weyl Fermions.  The covariant derivatives
are:
\beqra
D_\mu\psi &=&\left(\pa_\mu+ig W_\mu+\frac{2i}{3}\,g't_3\hat
B_\mu\right)\psi\nonumber \\[6pt]
D_\mu\psi &=&\left(\pa_\mu- ig'\left(1+\frac{2t_3}{3}\right)B_\mu\right)
\bar\psi
\eeqra

\end{document}